\documentclass[]{spie}  %>>> use for US letter paper
\usepackage[]{graphicx}

\usepackage{xcolor}

\usepackage{amssymb}
\usepackage{amsmath}
\usepackage{amscd}
\usepackage{bm}

\newcommand{\pll}{\parallel}

\newcommand{\rmd}{{\rm d}}
\newcommand{\rmi}{{\rm i}}

\newcommand{\de}{\delta}

\newcommand{\ignore}[1]{\relax}

\newcommand{\tDtwo}{\tau_{\rm D2}}

\newcommand{\tEo}{\tau_{\rm E}^{\rm op}}

\newcommand{\Nm}{N_m}
\newcommand{\Nmo}{N_{m_0}}
\newcommand{\Nmp}{N_{m'}}
\newcommand{\rhom}{\rho_m}
\newcommand{\rhomo}{\rho_{m_0}}
\newcommand{\rhomp}{\rho_{m'}}

\title{Shot-noise of quantum chaotic systems in the classical limit}

\author{Robert S. Whitney
\skiplinehalf
Institut Laue-Langevin,
6, rue Jules Horowitz, BP 156,
38042 Grenoble Cedex 9.
France.
\skiplinehalf
24 April 2007 (updated 5 Oct.\ 2007 \& erratum added for Eq. (16) in Dec.\ 2020)}

\begin{document} 
  \maketitle 

%%%%%%%%%%%%%%%%%%%%%%%%%%%%%%%%%%%%%%%%%%%%%%%%%%%%%%%%%%%%% 
\begin{abstract}
Semiclassical 
methods can now explain many mesoscopic effects
(shot-noise, conductance fluctuations, etc)
in clean chaotic systems, such as chaotic quantum dots.  
In the deep classical limit
(wavelength much less than system size) the Ehrenfest time
(the time for a wavepacket to spread to a classical size)
plays a crucial role, and random matrix theory (RMT) 
{\it ceases to apply} to the transport properties of open chaotic systems.

Here we summarize some of our recent 
results for shot-noise (intrinsically quantum noise in the current through
the system) in this deep classical limit.
For systems with perfect coupling to the leads, we 
use a phase-space basis on the leads to show that the transmission eigenvalues
are all 0 or 1 --- so transmission is noiseless 
[Whitney-Jacquod, Phys.~Rev.~Lett.~{\bf 94}, 116801 (2005), 
Jacquod-Whitney, Phys.~Rev.~B {\bf 73}, 195115 (2006)].
For systems with tunnel-barriers on the leads we 
use trajectory-based semiclassics to extract universal (but non-RMT)
shot-noise results for the classical regime [Whitney, 
Phys.~Rev.~B {\bf 75}, 235404 (2007)].

\end{abstract}


%%%%%%%%%%%%%%%%%%%%%%%%%%%%%%%%%%%%%%%%%%%%%%%%%%%%%%%%%%%%%
\section{INTRODUCTION}
\label{sect:intro} 
 
In recent years it has been possible to make quantum dots clean enough
that the electrons have a mean free path significantly longer than 
the size of the potential that confines them\cite{Revdot1,marcus}.
The electrons move ballistically in such a dot, 
in a manner strongly related to the {\it classical} dynamics 
associated with the dot's confining potential.
It has long been observed that
when this classical motion is {\it chaotic}, 
the properties of a closed quantum dot (such as 
energy-level statistics close to
the Fermi surface) are well-captured by random matrix theory (RMT)
\cite{BGS,Haake-levelstat}.
However for open quantum systems it has become increasingly clear that the
situation is very different, see Fig.~\ref{fig:regimes}.
The cross-over to non-RMT behaviour happens when 
an {\it Ehrenfest time}
becomes of order (or greater than) the dwell time
(the typical time the 
particles spend in the chaotic dot) \cite{Ale96,agam}.
The Ehrenfest times are the times for a wavepacket to spread
(under the classical dynamics) from a size of order a wavelength
to a classical scale (i.e.~system size, lead widths, etc).

In this article we give a brief overview
of our recent results on the nature of quantum noise in the new ``classical''
regime, where Ehrenfest times are much greater than the dwell time.  
For more detailed information on the calculations, 
we refer the reader to the works cited in each section.
Similarly those works contain results and discussions on the 
cross-over from the random matrix regime to the classical regime, 
which we omit here.

\subsection{Quantum dots: a laboratory for quantum chaos}

One of the most fundamental questions in quantum mechanics is
how the every-day world that we experience emerges from  
a sea of particles obeying quantum mechanics.
Since it is well known that many things in the everyday world are chaotic
(the weather, etc), we should try to understand how classical 
chaos emerges from quantum mechanics \cite{Haake-book}.
There are two things one can do to take the classical limit of a 
quantum system.
\begin{itemize}
\item
{\bf Vanishing wavelength:} Taking the ratio of the particle's wavelength to
all other lengthscales to zero.  Usually this means the wavelength becomes much less than the detector size, making quantum 
interference effects hard to observe.
\item
{\bf Decoherence:} The particles being studied
often interact with other particles in their environment. This can lead 
to the loss of phase information
and the suppression of quantum interference effects.
\end{itemize}

\begin{figure}
   \begin{center}
   \begin{tabular}{c}
   \includegraphics[width=14cm]{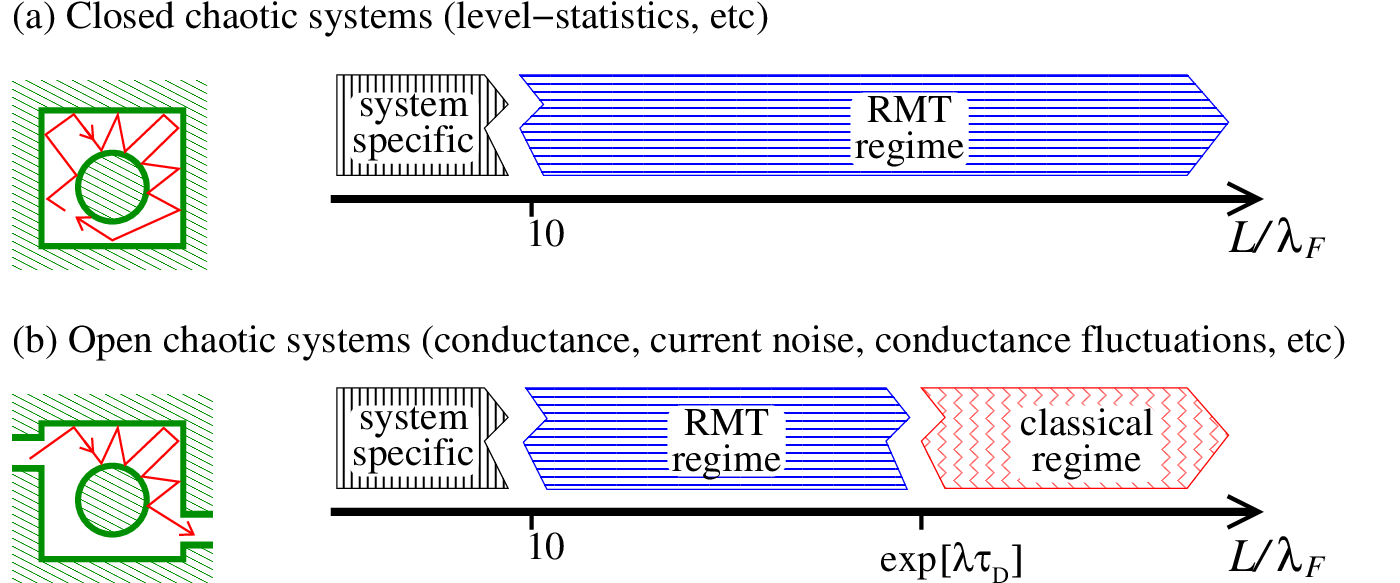}
   \end{tabular}
   \end{center}
   \caption[] 
   { \label{fig:regimes} 
Regimes for closed and open chaotic systems as one goes to the classical limit,
when the ratio of system size, $L$, to wavelength, $\lambda_{\rm F}$, 
goes to infinity.  
On the left we have cartoons of classical motion in 
a chaotic system (Sinai billiard) which is (a) closed and (b) open.
Note that we assume that the shape of the system is 
unchanged as we take  $L/\lambda_{\rm F}\to \infty$, thus we assume the
ratio of lead width, $W$, to system size remains constant.
For an open system the cross-over between the {\it RMT} and {\it classical} 
regimes happens when the Ehrenfest time $\tau_{\rm E}^{\rm cl} 
= \lambda^{-1} \ln[L/\lambda_{\rm F}]$ grows to become larger than 
the dwell time, $\tau_{\rm D}$.}
   \end{figure}

To get experimental insight into quantum chaos, one must take a system 
whose shape would induce chaos in classical particles,
and insert a quantum particle whose wavelength 
is much smaller than the system size,
but {\it not} immeasurably smaller.
Micron-sized (i.e. big) quantum dots are ideal for this,
where the wavelength is given by the Fermi surface and is
typically a few nanometres.
It is crucial that the dots are extremely clean,
since impurities typically have a size of order the electron wavelength, 
and so
cause highly quantum (s-wave) scattering, independent of the ratio
of $L$ to $\lambda_{\rm F}$.
By varying the dot's temperature, one can control 
the amount of decoherence \cite{footnote:temperature}.
Thus quantum dots are ideal laboratories for answering the basic questions of 
quantum chaos.
The first experimental observation of the cross-over 
between the RMT and classical regimes (Fig.~\ref{fig:regimes}b) 
was made for
the a measure of the ratio of shot-noise to current (the Fano factor)
in such a device\cite{eugene2}.

In different situations, the relative importance of the two classical limits
given above (vanishing wavelength and decoherence) are different.
Decoherence plays a crucial role in 
weak-localization\cite{Ale96,petitjean06,wjp07} 
and conductance fluctuations\cite{Brouwer-quasi-ucf} 
in quantum chaotic dots.
However shot-noise is insensitive to 
decoherence\cite{Been-review,vanLangen97,wjp07}, 
so in this article we can neglect decoherence effects entirely.

%-------------
%-------------

\subsection{Ehrenfest times}
Ehrenfest times are the time-scales on which quantum effects 
start to become relevant in the evolution of a wavepacket. 
They have acquired this name because
Ehrenfest's theorem (that
quantum wavepackets evolve in the same way as
a classical probability distributions) is only valid up to these timescales.

We consider a chaotic cavity  of
size $L$ and Lyapunov exponent $\lambda$
which is connected to leads of width $W$; 
where  $L,W$ are all much larger than the Fermi wavelength,
$\lambda_{\rm F}=\hbar/p_{\rm F}$.
There are  
Ehrenfest times 
associated with each classical scale\cite{Vavilov,Scho05}; 
\begin{eqnarray}
\tau_{\rm E}^{\rm cl} 
= \lambda^{-1} \ln \big[ L/\lambda_{\rm F}\big] 
\qquad \hbox{ and } \qquad  
\tau_{\rm E}^{\rm op} 
= \lambda^{-1} \ln \big[(L/\lambda_{\rm F}) \times (W/L)^2\big] .
\end{eqnarray}
The former we call the {\it closed cavity} Ehrenfest time
as it is the only such timescale for a closed chaotic system.
The latter we call the {\it open cavity} Ehrenfest time
as it is associated with the presence of leads 
(although both $\tau_{\rm E}^{\rm op}$ and $\tau_{\rm E}^{\rm cl}$
are relevant in open systems).
These scales can be derived as follows.
We assume the cavity is a two-dimensional hyperbolic chaotic system. Then the 
Poincar\'e surface of section perpendicular to any trajectory is 
a two-dimensional phase space ($r_\perp,p_\perp$), which we  
make dimensionless by writing distances in units of 
$L$ and momenta in units of $p_{\rm F}$.
Then the Liouvillian flow on the Poincar\'e surface of section  
stretches exponentially, with rate $\lambda$  in the {\it unstable} 
direction,
while contracting exponentially in the {\it stable} direction.
The Ehrenfest times are then given by 
$\lambda^{-1} \ln [\hbar_{\rm eff}^{-1} X^2]$ 
where $X$ is
a dimensionless system lengthscale, $W/L$ or $1$, 
and $\hbar_{\rm eff}=\lambda_{\rm F}/L$.
This is the time for a wavepacket with width $X$ 
in the {\it stable} direction (and hence $\hbar_{\rm eff}/X$ in the 
{\it unstable} direction) to spread under the Liouvillian flow 
to width $X$ in the {\it unstable} direction. 

\subsection{Shot-noise and transmission eigenvalues}

In this article we discuss the zero-frequency shot-noise
power, $S$, for a quantum chaotic systems. 
This is the intrinsically quantum part of the fluctuations of 
a non-equilibrium electronic current
and it contains information that cannot be obtained through
conductance measurements. 
We give our results in terms of the Fano factor $F=S/S_{\rm p}$,
which is the ratio of $S$ to the Poissonian noise, 
$S_{\rm p}=2 e \langle I \rangle$, that a current 
flow of uncorrelated particles would generate.
As such, the Fano factor is a measure of the 
ratio of the noise to the average current.
The scattering theory of
transport\cite{Blanter-review} gives
\begin{eqnarray}
F&=& 
{{\rm Tr}[{\mathbb S}_{mm_0}^\dagger {\mathbb S}_{mm_0}]
-{\rm Tr}[{\mathbb S}_{mm_0}^\dagger {\mathbb S}_{mm_0}
{\mathbb S}_{mm_0}^\dagger {\mathbb S}_{mm_0}] 
\over 
{\rm Tr}[{\mathbb S}_{mm_0}^\dagger {\mathbb S}_{mm_0}] },
\label{eq:Fano-definition}
\end{eqnarray}
where ${\mathbb S}_{mm_0}$ is a matrix made up of those
elements of the scattering matrix, ${\mathbb S}$, 
which correspond to scattering
from an ingoing mode on lead $m_0$ to an outgoing mode on lead $m$.
If we can diagonalize ${\mathbb S}_{mm_0}$, then it is trivial 
to extract the Fano factor, since it is given by the following function of
the eigenvalues, $\{t_i\}$, of ${\mathbb S}_{mm_0}$;
\begin{eqnarray}
F&=&  {\sum_i t_i^*t_i (1 - t_i^*t_i)
\over \sum_i t_i^*t_i}. \quad
\label{eq:Fano-eigenvalues}
\end{eqnarray}
Crucially this means that all modes with an eigenvalue, $t_i$,
which has a magnitude equal to 0 or 1, will {\it not} contribute to the noise. 

\section{PS-basis: diagonalizing most of the scattering matrix
\cite{WJ2004,JW2005}}

%%%%%%%%%%%%%%%%%%%%%%%%%%%%%%%%%%%%%%%%%%%%
%% Figure 1 : wavepacket
%%%%%%%%%%%%%%%%%%%%%%%%%%%%%%%%%%%%%%%%%%%%
\begin{figure}   
\begin{center}
   \begin{tabular}{c}
   \includegraphics[width=7cm]{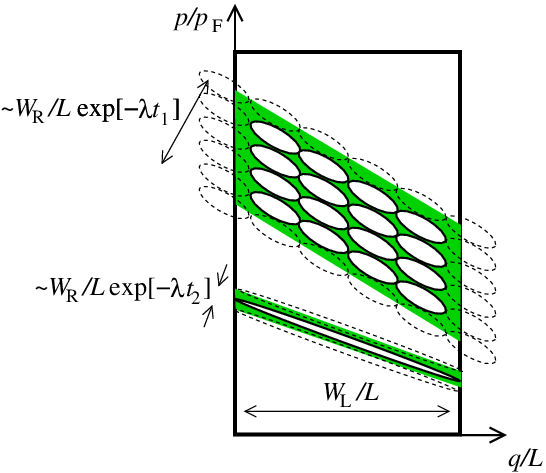}
   \hskip 1cm \includegraphics[width=6cm]{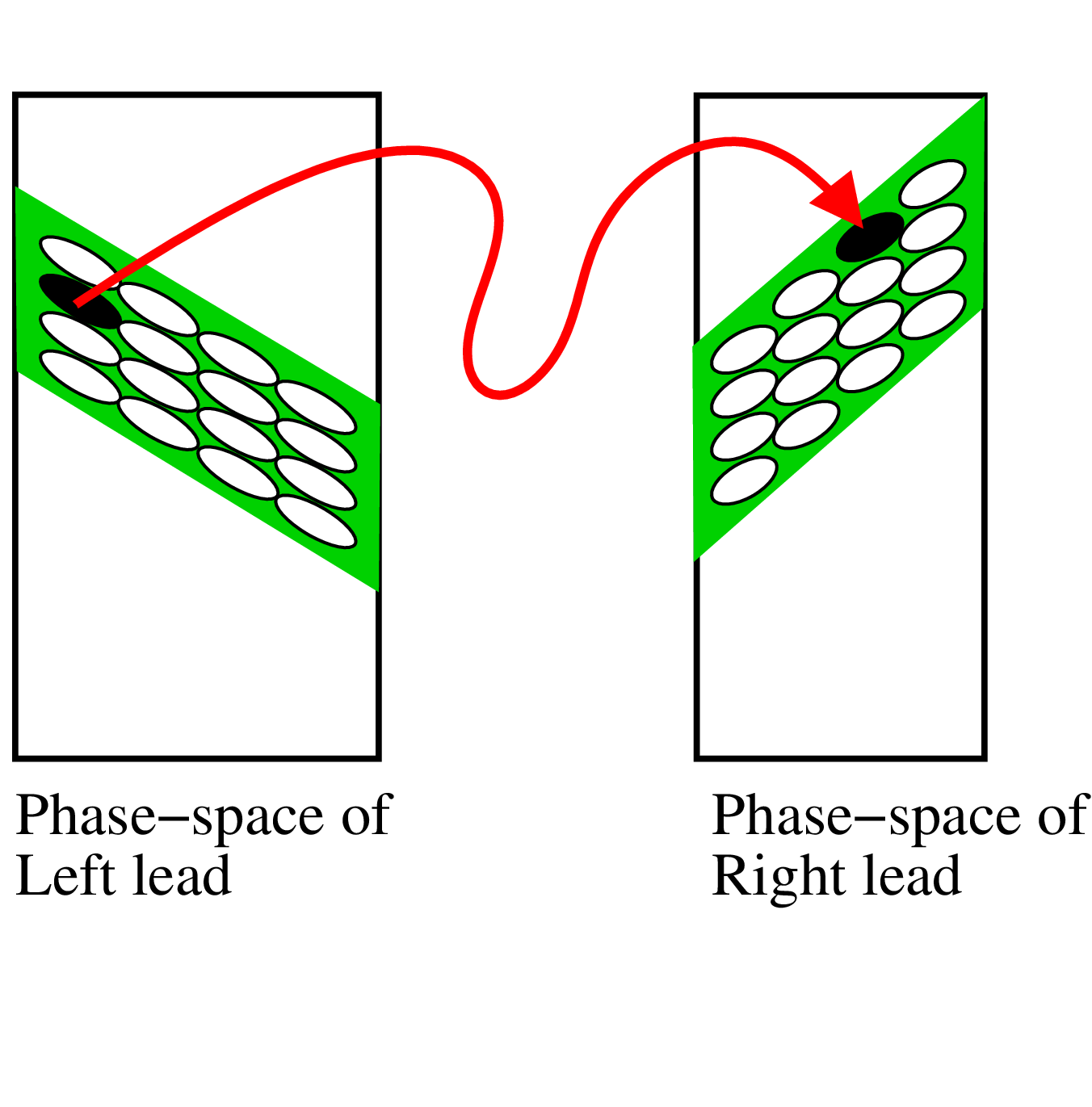}
   \end{tabular}
   \end{center}
   \caption[] 
   { \label{fig:bands} 
On the left we sketch two bands (in green) on the phase-space of the Left lead.
PS-states (ellipses) are super-imposed on these bands.
The phase-space is dimensionless, with all lengths  and momenta in units of
$L$ and $p_{\rm F}$, so each ellipse has an area of 
$\hbar_{\rm eff} = h/(p_{\rm F}L)$. 
The lattice of PS-states has been stretched/rotated 
to maximize the number of PS-states in each band (solid-edged ellipses) 
while minimizing the number partially in each band (dashed-edged ellipses).  
Thus the PS-states have the same aspect ratio as the band.
On the right we show the one-to-one correspondence between incoming
modes in the band on the Left lead, and outgoing modes in the bands
on the Right lead.
}
\end{figure}
%%%%%%%%%%%%%%%%%%%%%%%%%%%%%%%%%%%%%%%%%%%%%%%%%%%

\subsection{Bands in the classical phase-space}
The finiteness of $\tau_{\rm D}$ (the dwell time for trajectories in the 
cavity) means that classical trajectories injected into a cavity 
are naturally grouped into transmission and 
reflection bands \cite{Wirtz99,silvestrov} in phase-space (PS), 
despite the ergodicity of 
the associated closed cavity. 
Each band on the PS cross-section of the L lead (see Fig.~\ref{fig:bands})
consists of a group of classical paths which exit through the same lead 
after the same number of bounces, $\tau$,
(having followed similar paths through the cavity).
Because of the chaotic classical dynamics, bands with longer 
escape times are narrower, having a width (and hence a PS area) 
scaling like $\propto \exp[-\lambda \tau]$.
The open-cavity Ehrenfest time, $\tEo$, 
is the time at which this area becomes smaller
than $\hbar$.
Thus for times {\it shorter} 
than this, $\tau < \tau_{\rm E}^{\rm op}$, 
a band can carry one (or more)
{\it orthogonal} quantum wavepackets.
We argue below that these can be associated with PS-states 
(lead modes in the phase-space basis) which behave 
{\it classically}.
Hence the number of transmitting {\it classical} PS-states 
is given by the area of the L lead's phase-space which couples to transmitting 
trajectories with $\tau < \tau_{\rm E}^{\rm op}$.
The total number of classical modes in the L lead is the sum of this 
and the bands which reflect in a time $\tau < \tau_{\rm E}^{\rm op}$; 
\begin{eqnarray}
N^{\rm cl}_{\rm L}
&=&  N_{\rm L} (1-e^{-\tau_{\rm E}^{\rm op}/\tau_{\rm D}})
\end{eqnarray}
where we assume the leads have similar enough width that
the Ehrenfest time for transmission and reflection are almost the same
\cite{JW2005}.
All other modes of the L lead sit over many 
transmission or reflection bands with $\tau > \tau_{\rm E}^{\rm op}$,
and so they are {\it quantum} PS-states;
thus 
$ N^{\rm qm}_{\rm L}= N_{\rm L} e^{-\tau_{\rm E}^{\rm op}/\tau_{\rm D}}$.
We can do the same for the phase-space of the R lead by 
replacing L with R throughout.

\subsection{Scattering matrix in the phase-space basis}

We now summarize the construction of the PS-basis;
a basis made of states that are all localised in phase-space
(for details see Ref.\cite{JW2005}).
We cover all phase-space bands with areas bigger 
than $2\pi\hbar$ with a lattice of 
PS-states of the form shown in Fig.~\ref{fig:bands}.
The lattice is stretched and rotated to optimally cover each band.
We can use wavelet analysis to ensure that 
the lattice of states covering each such band is {\it complete} and 
{\it orthonormal} (within each band).
We choose the lattice's position on each band such that each ingoing
PS-state evolves under the cavity dynamics to exit as exactly one outgoing
PS-state. In this construction, each basis states exits 
at a time less that $\tau_{\rm E}^{\rm op}$.  
It behaves completely {\it deterministically}, 
i.e.~like a classical particle.  It
exits as a single wavepacket at a single time through a single lead,
completely hiding its quantum nature.

We complete the basis by covering the remaining phase-space 
(covered in classical bands with phase-space area less than $2\pi\hbar$) 
in whatever manner is required to complete the orthonormal basis.
The basis is already complete
on the bands with area larger than $2\pi\hbar$, so each
remaining PS-states must sit on many bands in the classical phase-space which
exit at many different times through different leads.  
Thus these PS-basis states 
exhibit strongly quantum behaviour, however for $\tEo \gg \tau_{\rm D}$ 
the proportion
of such quantum states vanishes.

The basis of lead modes and the PS-basis are related to each other by a 
{\it unitary} transformation, because both bases are complete and orthonormal.
Such a transformation leaves the eigenvalues of the 
scattering matrix, ${\mathbb S}$, unchanged.
As such the transformation should not change any of the transport properties
of the system (they all involve only traces of products of 
${\mathbb S}_{mm_0}^\dagger {\mathbb S}_{mm_0}$). 
The scattering matrix in the PS-basis is,
\begin{eqnarray}\label{eq:splitting}
{\mathbb S} = {\mathbb S}_{\rm cl} \oplus {\mathbb S}_{\rm qm} 
=\left( \begin{array}{cc} 
{\mathbb S}_{\rm cl} & 0 \\ 0 & {\mathbb S}_{\rm qm}
\end{array}\right)
\end{eqnarray}
The one-to-one correspondence between ingoing and out-going modes
on each band means that 
${\mathbb S}_{\rm cl}$ has only one non-zero element in
each row and column.  If for a system with two leads (L and R), 
we re-order the labels of the modes on L and R, we can write
\cite{WJ2004,JW2005}
\begin{eqnarray}
{\mathbb S}_{\rm cl} 
\ \equiv\ 
\left( \begin{array}{cc} 
{\bf r}_{\rm cl} & {\bf t}'_{\rm cl} \\ {\bf t}_{\rm cl} & {\bf r}'_{\rm cl}
\end{array}\right)
\qquad \hbox{ with }  \qquad
{\bf t}_{\rm cl} 
= \left( \begin{array}{cc} 
\tilde{\bf t}_{\rm cl} & 0 \\ 0 & 0
\end{array} \right)
\qquad \hbox{ and } \qquad 
{\bf r}_{\rm cl} 
= \left( \begin{array}{cc} 
0 & 0 \\
0 & \tilde{\bf r}_{\rm cl}
\end{array} \right).
\end{eqnarray}
The matrices 
$\tilde{\bf t}_{\rm cl}$ and 
$\tilde{\bf t}'_{\rm cl}$ are $n\times n$,
where 
$n= [N_{\rm L} N_{\rm R}/(N_{\rm L}+ N_{\rm R})] 
\exp [-\tau_{\rm E}^{\rm op}/\tau_{\rm D}]$ 
is the number of {\it classical transmission modes}.
The matrix $\tilde{\bf r}_{\rm cl}$ is
$(N_L^{\rm cl}-n)\times(N_L^{\rm cl}-n)$ 
and 
$\tilde{\bf r}'_{\rm cl}$ is 
$(N_R^{\rm cl}-n)\times(N_R^{\rm cl}-n)$. 
The matrix $\tilde{\bf t}_{\rm cl}$ is diagonal with elements given by
$\tilde{t}_{ij} = e^{{\rm i} \Phi_i}\delta_{ij}$
The matrix $\tilde{\bf r}_{\rm cl}$ has a slightly more complicated structure,
but it still has exactly one non-zero element in each row and each column.
Thus we have diagonalized  
$N_{\rm L}^{\rm cl}$ of the modes of ${\mathbb S}$. 
It has $n$ modes with eigenvalues obeying $|t_i|=1$
and $N_{\rm L}^{\rm cl}-n$ modes with eigenvalue $t_i=0$.  
From Eq.~(\ref{eq:Fano-eigenvalues}), 
we see that all these modes are noiseless. 
In the classical limit the proportion of such classical (noiseless)
modes goes to one \cite{vanhouten}.
The remaining modes remain numerous, but their proportion goes to 
zero. They are quantum in nature and are unitary {\it within} their own
subspace, ${\cal S}_{\rm qm}$. 

This gives a microscopic proof of an earlier prediction 
that the transmission eigenvalues
behave as if the system splits into two systems in parallel
(one classical, one quantum) \cite{silvestrov};
however it does not say anything about whether
the quantum system has RMT behaviour or not.
As the classical modes are {\it noiseless}, 
all noise is generated by the quantum modes. 
Thus we can expect the Fano factor $\propto$ 
(2nd moment of noise/average current)
to scale like $\exp [-\tau_{\rm E}^{\rm op}/\tau_{\rm D}]$,
vanishing as $\hbar \to 0$.
This fits numerical and experimental \cite{eugene2} observations and 
has agreement with the earlier microscopic theory 
\cite{agam}.  

After performing this phase-space analysis, we were able to apply a more
traditional (real-space) 
semiclassical approach to the Fano factor\cite{wj2005-fano}
(thereby reproducing Ref.~\cite{agam}), this was then extended 
to the third-moment of the noise\cite{Brouwer-noise}.
These works show that the
quantum modes do indeed fit RMT (for the reduced part of the scattering 
matrix that they inhabit) up to at least the third moment of the noise.
However we caution the reader that this {\it effective-RMT 
conjecture}\cite{silvestrov} does not work for other quantities, such
as weak-localization~\cite{Brouwer-quasi-wl}.

\section{Shot-noise with tunnel-barriers\cite{W2006-TUNNEL}}

We now consider a situation where the leads are not
perfectly coupled to the chaotic system.
Instead the particles must tunnel through a barrier to enter or leave the 
system.  We consider the limit where the ratio of $L,W$ to $\lambda_{\rm F}$
goes to infinity while the tunnelling probability, $\rho$, remains constant.
This is {\it not} the standard classical limit,
because it requires that the thickness of the barriers scale with 
$\lambda_{\rm F}$ not $L$.
However if this thickness were to scale with $L$, then all
barriers would become impenetrable in the classical limit, 
and there would be no interesting physics to investigate!
With tunnel-barriers, 
the phase-space splitting method no longer works;
the barriers mix the PS-basis states
because the wavepacket is part transmitted and part reflected
 each time the wavepacket hits a barrier.
Thus instead we added tunnelling effects to 
the trajectory-based semiclassical
method\cite{W2006-TUNNEL} 
previously used for noise without barriers\cite{wj2005-fano,Haake-fano}
(see also work on quantum graphs\cite{Schanz-fano}).

%%%%%%%%%%%%%%%%%%%%%%%%%%%%%%%%%%%%%%%
\begin{figure}
\begin{center}
% \hspace{-1cm}
\resizebox{15cm}{!}{\includegraphics{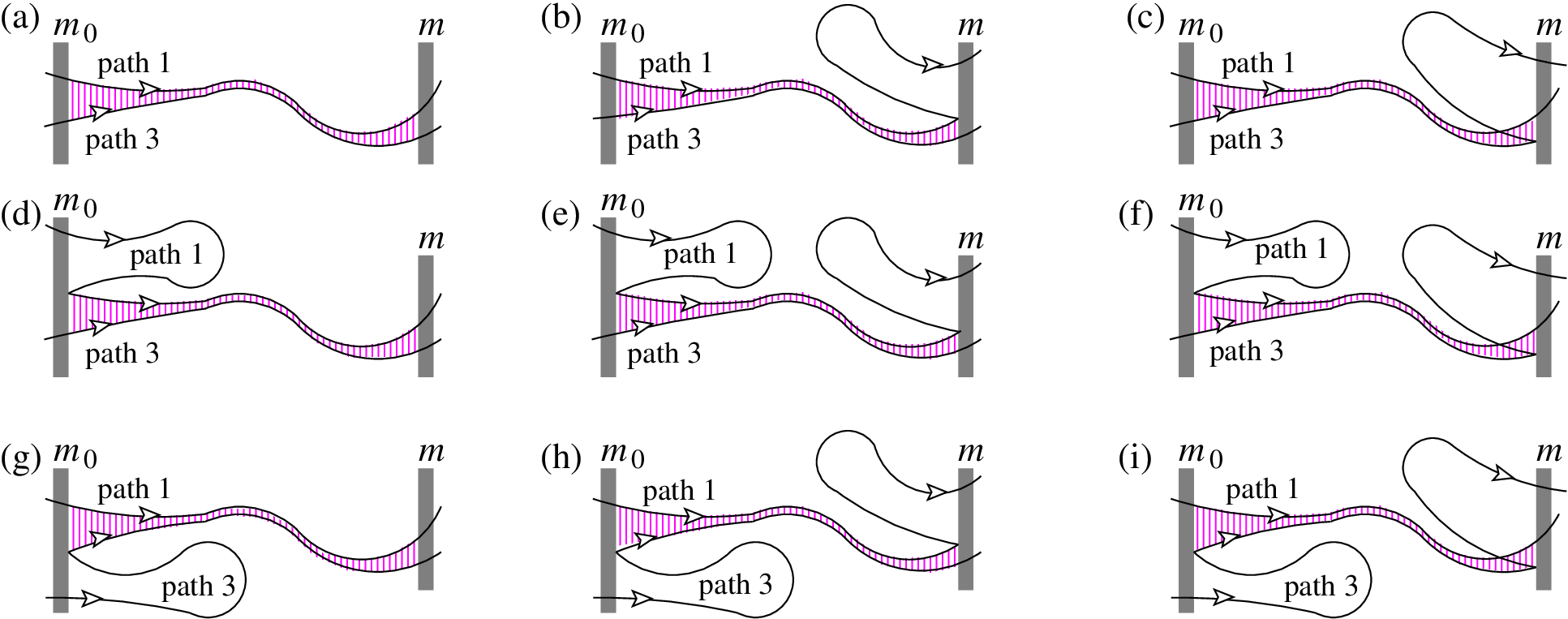}}
\caption{\label{fig:noise-trans1}
Calculating the shot-noise in the presence of tunnel-barriers.
The set of contributions to 
${\rm tr}[{\mathbb S}_{mm_0}^\dagger {\mathbb S}_{mm_0}
{\mathbb S}_{mm_0}^\dagger {\mathbb S}_{mm_0}]$
which do not vanish for infinite Ehrenfest time. 
Here we show only the tunnel-barriers on leads $m_0$ and $m$ as shaded 
rectangles, a path which crosses the barrier on lead $m$ has succeeded
in tunnelling out of the cavity into the lead.  
The contributions are made up of four classical paths,
here we show only two of the paths (1 and 3). The other two paths (2 and 4)
look the same as the paths shown, {\it except} that they cross
at the centre of the correlated region (indicated by the vertical 
cross-hatching).  
Thus path 4 is paired with path 1 at lead $m_0$ but
paired with path 3 at lead $m$ (and vice-versa for path 2).
The noise in these contribution is purely due to the stochastic nature 
of scattering at the tunnel-barriers, if the barriers were absent these
contributions would be noiseless.  
}
\end{center}
\end{figure}
%%%%%%%%%%%%%%%%%%%%%%%%%%%%%%%%%%%%%%%%%%

In the deep classical limit, $\tEo \to \infty$, 
all contributions (to lowest order in $1/N$) 
are listed in Fig.~\ref{fig:noise-trans1}. 
The contributions 
involve classical paths (path 1 and 3) which are paired 
(closer than $W$ with almost parallel momenta) in the cross-hatched region.
The encounter
is at the centre of this cross-hatched region,
it is shown in detail in Fig.~\ref{fig:encounter-at-tunnel}.
The distance between the paths at the encounter is of order 
$(\lambda_{\rm F}L)^{1/2}$, the reason for this will be sketched below.
The paths then diverge from each other 
as they move away from the encounter. 
However in the deep classical limit the time, $\tEo/2$, for paths to spread
from a distance apart $(\lambda_{\rm F}L)^{1/2}$ to a classical scale
become much larger than the dwell time.  Thus one
or both paths will escape {\it before} their flow under the cavity 
dynamics makes them become unpaired (diverge to a distance apart
greater than $W$).

The denominator and the first term in the numerator of 
Eq.~(\ref{eq:Fano-definition}) are equal to the dimensionless
Drude conductance
from lead $m_0$ to lead $m$, 
$g^{\rm D}_{\rm mm_0}
= {\rhomo\rhom\Nmo\Nm /\big(\sum_{m'} \rhomp \Nmp\big)}$.
To get this result one simply notes that there are $\Nmo$ incoming mode, each of which has a probability of $\rhomo$ to tunnel into the chaotic system,
and then a probability of $\rhom\Nm /\sum_{m'} \rhomp \Nmp$ of eventually 
escaping into the $m$th lead.
However to find the Fano factor we must also evaluate
${\rm Tr}[{\mathbb S}_{mm_0}^\dagger {\mathbb S}_{mm_0}
{\mathbb S}_{mm_0}^\dagger {\mathbb S}_{mm_0}]$. 
We can write this as the following
sum over four paths, 
\begin{eqnarray}\label{trt4}
{\rm Tr}[{\mathbb S}_{mm_0}^\dagger 
{\mathbb S}_{mm_0}{\mathbb S}_{mm_0}^\dagger {\mathbb S}_{mm_0}]
= 
{1\over (2\pi \hbar)^2}
\!\int_{\rm L} \! \! 
\rmd y_{01} \rmd y_{03} \int_{\rm R} \! \rmd y_1 \rmd y_3 
\sum_{\gamma1,\cdots \gamma4} 
A_{\gamma4}^*A_{\gamma3} A_{\gamma2}^*A_{\gamma1}
\exp [\rmi\delta S/\hbar] ,
\\
 \hbox{ where $\gamma1$ goes from $y_{01}$ to $y_1$,} \ \
\hbox{$\gamma2$ goes from $y_{03}$ to $y_1$,} \ \  
\hbox{$\gamma3$ goes from $y_{03}$ to $y_3$}, \ \ 
\hbox{$\gamma4$ goes from $y_{01}$ to $y_3$}, 
\end{eqnarray}
with $y_{01},y_{03}$ on lead $m_0$ and  $y_1,y_3$ on lead $m$.
The amplitude $A_\gamma$ is related the square-root of the 
stability of the path (its exact form is given in Ref.~\cite{W2006-TUNNEL}) 
and $\delta S = S_{\gamma1}-S_{\gamma2}+S_{\gamma3}-S_{\gamma4}$
(we have absorbed all Maslov indices into the actions
$S_{\gamma i}$).
The dominant contributions that survive averaging over energy or cavity shape
are those for which
the fluctuations of $\delta S/\hbar$ are minimal.  
Their paths are pairwise identical everywhere except in the
vicinity of encounters. Going through an encounter,
two of the four paths cross each other, while the other two
avoid the crossing. They remain in pairs, though the pairing switches,
e.g. from $(\gamma1;\gamma4)$ and $(\gamma2;\gamma3)$ to 
$(\gamma1;\gamma2)$ and $(\gamma3;\gamma4)$.
Thus in Fig.~\ref{fig:noise-trans1} we show only paths $\gamma1$ and $\gamma3$.
The action difference $\delta S$ is then given by the difference
between the paths close to the encounter,
as in the case without tunnel barriers, the integral over
all possible encounters is dominated by those where the paths $\gamma1$ and $\gamma3$ come within
$(\lambda_{\rm F}L)^{1/2}$ of each other\cite{Sieber01,Ric02}.
The paths are always close enough to their partner
that their stabilities are the same.
This stability of a classical path can then be related to the path's 
probability to go to a given point in phase-space.  Hence 
all contributions can be written in the form
\begin{eqnarray}\label{contribution}
D_i 
=
{1\over (2\pi \hbar)^2}
\!\int_{\rm L} \! \! \rmd {\bf Y}_{01} \; \rmd {\bf Y}_{03} 
\!\int_{\rm R} \! \! 
\rmd {\bf Y}_{1} \; \rmd {\bf Y}_{3} \!\int \! \! \rmd t_1 \; \rmd t_3 
\;
\langle P({\bf Y}_1,{\bf Y}_{01};t_1) \;  P({\bf Y}_3,{\bf Y}_{03};t_3) 
\rangle \; \exp [\rmi\delta S_{D_i}/\hbar] \,,
\end{eqnarray}
where the subscripts $1,3$ indicate paths 1 and 3 respectively. 
Here $P({\bf Y},{\bf Y}_{0};t) \de {\bf Y}\de t$ 
is the probability to go from
${\bf Y}_{0}=(y_0,p_{y0})$ to within $\de {\bf Y}=\de y \de p_y$
of ${\bf Y}=(y,p_{y})$ in a time within $\de t$ of $t$.
For an individual system, this has a $\de$-function on each classical
path, however its average over energy or system shape is a smooth function.
Here we have to average over a pair of such probabilities
in situations in which the paths start or finish close to each other in 
phase-space.
This has the subtlty that 
when the paths are close to each other
their escape probabilities are highly correlated (if one path hits
a lead when the other is within $W$ of it, the probability that the second
path hits the lead is close to one).
Ref.~\cite{W2006-TUNNEL} discusses
the details of how to evaluate such probabilities.

\subsection{Evaluating the contributions.}
\label{sect:D1}

%%%%%%%%%%%%%%%%%%%%%%%%%%%%%%%%%%%%%%%
\begin{figure}
\begin{center}
% \hspace{-1cm}
\resizebox{16cm}{!}{\includegraphics{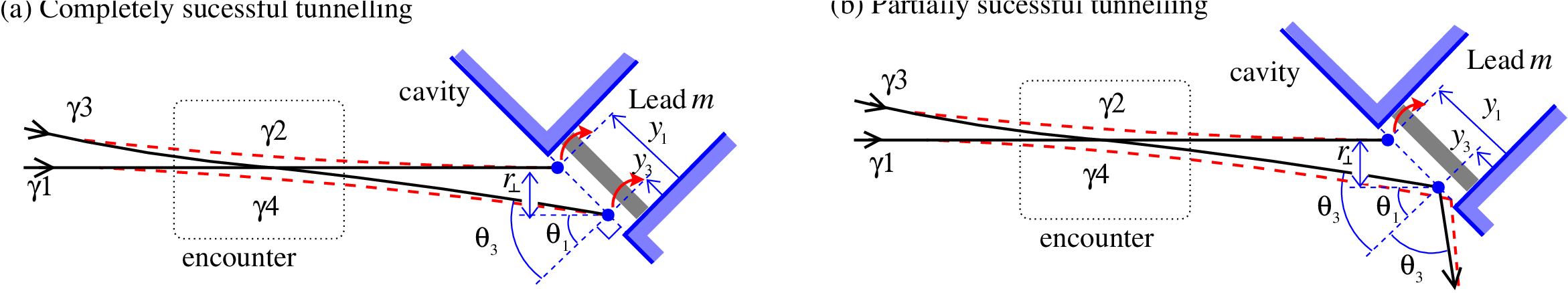}}
\caption{\label{fig:encounter-at-tunnel}
Details of the paths in Fig.~\ref{fig:noise-trans1}.
Path $\gamma 1$ (solid black line)
hits the cross-section of lead $m$ at position
$y_1$ with momentum angle $\theta_1$, while path $\gamma3$
hits the lead at $(y_3,\theta_3)$.
One path (in this case $\gamma1$) successfully escapes
while the other may (a) succeed in escaping or (b) fail to escape. 
All paths are drawn in the 
basis parallel and perpendicular to $\gamma1$ at escape, 
the initial position and momentum of path $\gamma3$ at the lead are
$r_{\perp} = (y_1-y_3)\cos \theta_1$, $r_{\pll} = (y_1-y_3)\sin \theta_1$  
and $p_{\perp} \simeq -p_{\rm F} (\theta_3-\theta_1)$.}
\end{center}
\end{figure}
%%%%%%%%%%%%%%%%%%%%%%%%%%%%%%%%%%%%%%%%%%

To evaluate all the contributions in Fig.~\ref{fig:noise-trans1}, we note
that the paths never become uncorrelated under the classical dynamics;
they only escape in an uncorrelated manner if one path tunnels while the 
other does not.
This is because we have taken $\tEo \to \infty$, so that paths 
with an encounter
take an infinite time to become uncorrelated (if the barriers are absent).
In this case the details of the
encounter are as given in Fig.~\ref{fig:encounter-at-tunnel}. 
Thus the action difference between 
the paths can be evaluated in a manner equivalent 
to coherent-backscattering with tunnel-barriers\cite{W2006-TUNNEL}, 
and
\begin{eqnarray}
\de S_{D_i}
&=& (p_{\perp}+m\lambda r_{\perp})r_{\perp}\,.
\label{eq:deltaS-double-success}
\end{eqnarray} 
for contributions of the form in
Fig.~\ref{fig:encounter-at-tunnel}a.
For contributions
of the form in Fig.~\ref{fig:encounter-at-tunnel}b,
the action difference is almost the same
(the difference has no effect on the integrals\cite{W2006-TUNNEL}) 
so we can use Eq.~(\ref{eq:deltaS-double-success}) there as well.

For the contribution in Fig.~\ref{fig:noise-trans1}a,
the paths paired when they hit lead $m$ were also paired
at lead $m_0$, thus the length of the paired region (cross-hatched in
Fig.~\ref{fig:noise-trans1}) must be less than
$T'_W$. The time, $T'_W$, 
is the time-difference between the paths 
differing by $(r_{\perp},p_{\perp})$  
and the {\it earlier} time when they would have been $W$ apart
(if the leads were absent).
We find that
\begin{eqnarray}
\int_m \! \rmd {\bf Y}_1  \rmd {\bf Y}_3
\int_0^{T'_W} \rmd t_1 \rmd t_3
\langle P({\bf Y}_1,{\bf Y}_{01};t_1) P({\bf Y}_3,{\bf Y}_{03};t_3) \rangle_{\rm 1a} 
% \nonumber \\
&=& {\rhom^2 \Nm p_{\rm F}^2 \cos \theta_{01} \cos \theta_{03} 
\over \sum_{m'} \rhomp (2-\rhomp)\Nmp} 
% \nonumber \\
% & & \times
(1-\exp[-T'_W/\tDtwo]). \qquad \ 
\label{eq:prob-for-D_1a}
\end{eqnarray}
where $\tDtwo$ is the survival time for paths which stay extremely close
to each other \cite{W2006-TUNNEL}.
The integral over $(r_{\perp},p_{\perp})$  is dominated by 
$r_{\perp}-(m\lambda)^{-1}p_{\perp} \sim (\lambda_{\rm F}L)^{1/2}$,
as a result the time $T'_W \to \infty$ in the classical limit,
so we can neglect
any terms of the form $\exp[-T'_W/\tDtwo]$.
The denominator comes from the fact we are considering the survival
probability for a pair of paths; the probability that 
the pair is destroyed
by one or both paths escaping into a lead during the time $t$ to $t+ \de t$
is $P_2(t) \times \de t/\tDtwo$.
We insert Eq.~(\ref{eq:prob-for-D_1a}) into Eq.~(\ref{contribution}),
then integrate over all possible $y_{03}$ and $p_{03}$.
We find the contribution to 
${\rm Tr}[{\mathbb S}_{mm_0}^\dagger {\mathbb S}_{mm_0}
{\mathbb S}_{mm_0}^\dagger {\mathbb S}_{mm_0}]$ shown in 
Fig.~\ref{fig:noise-trans1}a is
\begin{eqnarray}
D_{\rm 1a}
&=& {\rhomo^2 \rhom^2 \Nmo \Nm (1-\exp[-\tEo/\tDtwo])
\over \sum_{m'} \rhomp (2-\rhomp)\Nmp}   \, .
\label{eq:D1a}
\end{eqnarray}

All contribution in Fig.~\ref{fig:noise-trans1} 
are very similar to $D_{\rm 1a}$.
One can see that $D_{\rm 1b}$ and $D_{\rm 1c}$ 
are like  $D_{\rm 1a}$ with the exception that
a path is reflected off lead $m$ and then returns to lead $m$.
After reflection that path evolves alone in the cavity.
Thus each of  
these contributions is given by multiplying $D_{\rm 1a}$ by
\begin{eqnarray}
{1-\rhom \over \rhom }\times
{\rhom \Nm \over \sum_{m'}\rho_{m'}N_{m'}}  
&=& {(1-\rhom) \Nm \over \sum_{m'}\rho_{m'}N_{m'}},  
\label{eq:multiply-D_1a-by-this}
\end{eqnarray}
The same applies
for paths which enter the cavity from lead $m_0$ at different times,
in such a way that the paths form a pair, as in Fig.~\ref{fig:noise-trans1}d
(path 3 enters the cavity at a moment 
when path 1 is reflecting off barrier $m_0$, 
and both paths have similar momenta).  To see this we
reverse the direction of the paths, after which we have the 
situation discussed above with $m$ replaced by $m_0$.    
Summing all the contributions in Fig.~\ref{fig:noise-trans1} we get the
Fano factor in the deep classical limit ($\tEo \to \infty$)
\begin{eqnarray}
F 
&=& 1 -
{\rhomo\rhom  \sum_{m'} \rhomp \Nmp
\over \sum_{m'} \rhomp (2-\rhomp) \Nmp} 
\left(1+ {2(1-\rhomo)\Nmo \over \sum_{m'} \rhomp \Nmp} \right)
\left(1+ {2(1-\rhom)\Nm \over \sum_{m'} \rhomp \Nmp} \right).
\label{eq:Fano-classical-result}
\end{eqnarray}
If we kept $\tEo$ finite, the second term would contain a factor of
$(1-\exp[-\tEo/\tDtwo])$, however in this case we would not be able to
ignore other contributions (given in  Ref.~\cite{W2006-TUNNEL}
but neglected above) 
which go like $\exp[-\tEo/\tDtwo]$.

%%%%%%%%%%%%%%%%%%%%%%%%%%%%%%%%%%%%%%%
\begin{figure}
\begin{center}
% \hspace{-1cm}
\resizebox{12cm}{!}{\includegraphics{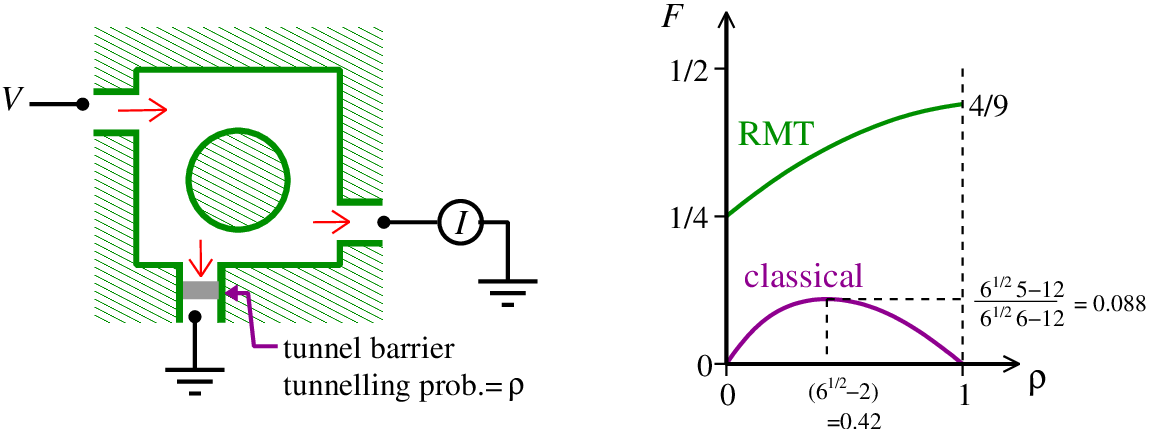}}
\caption{\label{fig:3lead}
Shot-noise measured in lead 2,
when current can also flow into another lead (lead 3).
For simplicity we assume all three leads have the same width.
On the right we give a cartoon of the Fano factor in the RMT and classical
regimes (the exact forms of the functions are given in 
Eq.~(\ref{eq:fano-3lead-class},\ref{eq:fano-3lead-rmt}).
}
\end{center}
\end{figure}
%%%%%%%%%%%%%%%%%%%%%%%%%%%%%%%%%%%%%%%%%%

\subsection{Shot-noise for a cavity with a third lead}

We now consider the special case (shown in Fig.~\ref{fig:3lead})
of a cavity with three leads,
the current is injected into one and detected at another
(neither of which have tunnel barriers), however the current can also
go through the tunnel-barrier into the third lead (where it escapes to earth without being not measured).
To keep the formulas as simple as possible we assume all leads have
the same width, 
so each has $N$ modes.
Then the Fano factor given in Eq.~(\ref{eq:Fano-classical-result}) 
reduces to
\begin{eqnarray}
F_{\rm 3leads}^{\tEo\to \infty} 
&=& {\rho(1-\rho) \over 2+ \rho(2-\rho)}
\label{eq:fano-3lead-class}
\end{eqnarray}
where $\rho$ is the transmission probability of the
barrier on the third lead.  
The Fano factor has a maximum
at $\rho=\sqrt{6}-1$, while it is zero (noiseless) 
when there is no tunnelling, 
i.e. when the barrier is either impenetrable ($\rho=0$) or absent ($\rho=1$). 
At the maximum the Fano factor is 
$(5\sqrt{6} -12)/(6\sqrt{6}-12)$ (see sketch of curve in Fig.~\ref{fig:3lead}).
This is completely different from the RMT result for the same system
(which is applicable for $\tEo \ll \tDtwo$),
\begin{eqnarray}
F_{\rm 3leads}^{\rm RMT} 
&=& {2 + 6\rho +4\rho^2+\rho^3 \over (2+\rho)^3}.
\label{eq:fano-3lead-rmt}
\end{eqnarray}
\vskip -2mm
\hskip 2cm \colorbox{pink}{
$\ $ \hskip 7mm\textsf{\textbf{ERRATUM (16 Dec 2020):} Eq.~(\ref{eq:fano-3lead-rmt})'s numerator should be $2 + 6\rho +3\rho^2+\rho^3$. \hskip 5mm$\ $}}
\vskip -2.5mm \hskip 2cm
\colorbox{pink}{
$\ $ \hskip 59.5mm\textsf{My thanks to Marcel Novaes for spotting this. \hskip 5mm $\ $}}

\vskip 2mm
The RMT result goes monotonically from the well-known two-lead result ($F=1/4$)
when $\rho=0$ to the three 3-lead result ($F=4/9$) when $\rho=1$.

\section{Conclusion: Universality of the classical regime}

We expect that almost any (hyperbolic) chaotic system (Sinai billiard,
stadium billiard, kicked-rotator maps, etc) will exhibit
the same average properties when coupled to leads.
Thus the results presented here for shot-noise in the deep classical limit
are {\it universal} without being given by random matrix theory (RMT).
The theory presented here is an ensemble of similar systems 
(with varying energy or system shape) rather than an individual system.
However for shot-noise we can estimate that the typical deviation of 
an individual system from the average results (calculated above)
vanishes in the classical regime (going like the inverse of the 
number of lead modes).

%%%%%%%%%%%%%%%%%%%%%%%%%%%%%%%%%%%%%%%%%%%%%%%%%%%%%%%%%%%%%

%%%%%%%%%%%%%%%%%%%%%%%%%%%%%
%%%%%%%%%%%%%%%%%%%%%%%%%%%%%
%%%%%%%%%%%%%%%%%%%%%%%%%%%%%


\begin{thebibliography}{1}
\bibitem{footnote:temperature}
Decoherence effects can be almost neglected
at the lowest experimentally-accessible temperatures, 
but they quickly grow one increases the temperature.
\end{thebibliography}
\end{document}